\documentclass[aps,prl,twocolumn,showpacs]{revtex4}

\usepackage{dcolumn}                    % Align table columns on decimal point
\usepackage{bm}                         % bold math
\usepackage{graphicx}

\newcommand {\ba} {\begin{eqnarray}}
\newcommand {\ea} {\end{eqnarray}}

\usepackage{times}

\begin{document}

\title{Anomalous scaling of the specific-heat jump $\Delta C$ vs. $T_c$ in the Fe-based
superconductors:\\ the $\pm$S-wave pairing state model}

\author{Yunkyu Bang$^{1}$ and G. R. Stewart$^{2}$}

\address{$^{1}$Department of Physics, Chonnam National University, Kwangju
500-757, Republic of Korea \\
$^{2}$ Physics Department, University of Florida, Gainesville, FL
32611-8440, USA}
%
%\email[$^{\S}$]{ykbang@chonnam.ac.kr}
%\email[$^{\dag}$]{stewart@phys.ufl.edu}

\begin{abstract}
The strong power law behavior of the specific heat jump $\Delta C$
vs. $T_c$ ($\Delta C/T_c \sim T_c ^{\alpha}, \alpha\approx 2$),
first observed by Bud'ko, Ni, and Canfield (BNC)\cite{BNC}, has
been confirmed with several families of the Fe-based
superconducting compounds with doping. We show here that this
anomalous non-BCS behavior is an intrinsic property of the
multiband superconducting state paired by a dominant interband
interaction ($V_{inter} > V_{intra}$) reflecting the relation
$\frac{\Delta_h}{\Delta_e} \sim \sqrt{\frac{N_e}{N_h}}$ near
$T_c$, as in the $\pm$S-wave pairing state. Then this $\Delta C$
vs. $T_c$ relation can continuously change from the ideal BNC
scaling to a considerable deviation at lower $T_c$ with a
moderate variation of the impurity scattering rate.
\end{abstract}


\date{\today}
\maketitle

{\it Introduction. ---} The specific heat (SH) jump $\Delta C$ is
the most well known thermodynamic signature of the second order
phase transition and hence contains the generic information of the
transition as well as the material specific information. For
example, the BCS theory of superconductivity predicts the
universal ratio $\Delta C /C_{el} |_{T=T_c} = 1.43$,  hence
$\Delta C /T_c =1.43 \gamma$ is a temperature independent constant
and tells us the material specific quantity $\gamma$, the
Sommerfeld coefficient of the normal state $\gamma = C_{el,n}/T$.
In view of this BCS prediction, $\Delta C /T_c =
const.$, the experimental observation by Bud'ko, Ni, and Canfield
(BNC)\cite{BNC}, $\Delta C /T_c \approx T_c^2 $ for a family of
doped Ba(Fe$_{1-x}$TM$_x$)$_2$As$_2$ compounds with TM=Co, Ni is a
very intriguing behavior and stimulated active investigations
both experimentally and theoretically. After the work
of Ref.\cite{BNC}, this so-called BNC scaling relation was expanded
with an increasing list of the iron pnictide and iron chalcogenide
(FePn/Ch) superconducting (SC) compounds\cite{JSKim2011,Hardy,Gofryk2010,Greg2012,AsP1,AsP2,Calfield_Na122,HHWen},
hence strengthens the speculation that some generic mechanism must exist behind this
unusual scaling behavior. However, the recent observation of a
strong deviation from the BNC scaling in a series of K-doped
Ba$_{1-x}$K$_x$Fe$_2$A$_2$ for $0.7 < x \leq 1$
\cite{Calfield_K122} is confusingly contrasted to the Na-doped
Ba$_{1-x}$Na$_x$Fe$_2$A$_2$ ($0.1 \leq x \leq
0.9$)\cite{Calfield_Na122} which displays an excellent BNC
scaling.

For the theoretical investigations, there are three
attempted explanations. Kogan\cite{Kogan2009} argued that strong
pair-breaking can cause $\Delta C /T_c \propto T_c ^2$. The
essence of this theory is a dimensional counting.
The free energy difference near $T_c$,
$\Delta F = F_{s} - F_{n}$, can be expanded in powers of
$\Delta^2$ ($\Delta$: the SC order parameter (OP)). In the BCS
theory, $\Delta F \propto -N(0)\frac{\Delta^4}{T_c ^2}$ \cite{AGD}.
Using the BCS result of
$\Delta^2(T)\sim T_c^2 (1 - \frac{T}{T_c})$, we get $\Delta C /T_c \propto
\frac{\partial^2 \Delta F}{\partial T^2} \sim N(0)$, the well
known BCS prediction. In the case of the strong pair-breaking
limit, $\Gamma_{\pi} \gg T_c$ ($\Gamma_{\pi}=$ pair-breaking rate),
considered by Kogan, $\Delta F \propto -
N(0)\frac{\Delta^4}{\Gamma_{\pi}^2}$ by a dimensional counting.
Substituting the same BCS behavior of $\Delta^2(T) \propto T_c^2
(1 - \frac{T}{T_c})$, we recover the Kogan's result $\Delta C /T_c
\sim N(0)\frac{T_c ^2}{\Gamma_{\pi}^2}$.
However, we believe that this result is the consequence
of an inconsistent approximation\cite{Kogan note}.
The theory of Vavilov {\it et al.}\cite{Vavilov2011} mainly
studied the coexistence region with magnetic order $M$ and SC
order $\Delta$. It is a plausible theory that the coexisting magnetic order over
the SC order can substantially reduce $\Delta C$, hence develops a
steep variation of $\Delta C$ vs. $T_c$. However this theory
didn't reveal a specific reason as to why $\Delta C /T_c$ follows
the BNC scaling  $\sim T_c ^{\alpha}$ with $\alpha \approx 2$.
Finally, Zannen\cite{Zaanen2009} attributed the behavior $\Delta C
\propto T_c ^3$ to the normal state electronic SH with the scaling form
$C_{elec}^{n} \propto T^3$ due to the critical fluctuations
near the quantum critical point (QCP). A problem of this theory is that there
is no evidence of $C_{elec}^{n} \propto T^3$ for a wide doping
range of the FePn/Ch superconductors.
All three theories mentioned above are single band theories and do not
particularly utilize the unique properties of the FePn/Ch superconductors.
In this paper, we propose a theory in which the
multi-band nature of the FePn/Ch superconductors is the root cause
for producing the BNC scaling behavior.

{\it Two Band model for the SH jump $\Delta C$. ---}
For a multi-band superconductor, the SH jump formula is
generalized as
\begin{equation}
\Delta C = \sum_{i=h,e} N_i(0) \Big( \frac{-d \Delta^2 _i}{d T} \Big)\Big|_{T_c}
\end{equation}
where the band index "$i$" counts the different bands
and we specify it as the hole and electron band typical in
the Fe-based superconductors. $N_{h,e}$ are the DOSs, and
$\Delta_{h,e}$ are the SC OPs of each band.
In the one band BCS superconductor, using $\Delta^2(T) \sim T_c^2
(1 - \frac{T}{T_c})$, the above equation gives $\Delta C /T_c
\propto N(0)=const$.
However, in the case of a multiband superconductor, Eq.(1)
can reveal more information for the pairing mechanism
as well as the pairing state.

At present the most widely accepted pairing state in the Fe-based
superconductors is the sign-changing S-wave state ($\pm$S-wave)
mediated by a dominant interband repulsive interaction
($V_{inter}> V_{intra}$)\cite{Mazin}. The essential physics of
this $\pm$S-wave state can be studied with the two coupled gap
equations\cite{Bang-imp}
\begin{eqnarray}
\Delta_h  &=&   -  \bigl[ V_{hh}N_h \chi_h  \bigr] \Delta_h
 -   \bigl[ V_{he} N_e \chi_e  \bigr]
\Delta_e ,
\\ \nonumber
\Delta_e  &=&   -  \bigl[ V_{ee} N_e \chi_e   \bigr] \Delta_e  -
\bigl[ V_{eh} N_h \chi_h  \bigr] \Delta_h ,
\end{eqnarray}
where the pair susceptibility at $T_c$ is defined as
\begin{equation}
\chi_{h,e}(T_c) = T_c \sum_{n} \int ^{\Lambda_{hi}} _{-\Lambda_{hi}} d\xi
\frac{1}{\omega_n ^2 +\xi^2} \approx \ln \Big[\frac{1.14 \Lambda_{hi}}{T_c}\Big],
\end{equation}
where $\omega_n =\pi T_c(2n+1)$ and $\Lambda_{hi}$ is a pairing energy cut-off.
The pairing potentials $V_{ab}$ ($a,b = h,e$) are all positive and further simplified in this paper as
$V_{he}=V_{eh}=V_{inter}$ and
$V_{hh}=V_{ee}=V_{intra}$ without loss of generality.

In the limit $V_{intra}/V_{inter} \rightarrow 0$, Eq.(1) can be
analytically solved and provides the interesting kinematic constraint
relation\cite{Bang-model}
\begin{eqnarray}
\frac{\Delta_h}{\Delta_e} \sim \sqrt{\frac{N_e}{N_h}} ~~~~{\rm as~~} T \rightarrow T_c ,
\end{eqnarray}
and the critical temperature is given by
\begin{equation}
T_{c} \approx 1.14 \Lambda_{hi} \exp{\big[-1/(V_{inter}\sqrt{ N_e  N_h})\big]}.
\end{equation}

For further modeling the calculation of the experimental data of
$\Delta C$ vs. $T_c$ for a Fe-122 compound with a series of
doping, we first notice that the undoped parent compound such as
BaFe$_2$As$_2$ is a compensated metal, hence has the same number
of electrons and holes, i.e. $n_h = n_e$. Therefore it is a
reasonable approximation for our model to take $N_h = N_e$ at no
doping and then the doping of holes (K, Na, etc.) or electrons
(Co, Ni, etc.) is simulated by varying $N_h$ or $N_e$ while
keeping $N_e + N_h =N_{tot} = const.$ For the rest of this paper,
it is convenient to use the normalized DOSs as
$\bar{N}_{h,e}=N_{h,e}/N_{tot}$ and $N_{tot}$ is combined to
define the dimensionless coupling constants as
$\bar{V}_{intra/inter}=N_{tot}\cdot V_{intra/inter}$.

Expanding the gap equations  Eq.(2) near $T_c$ and using Eq.(4), we obtain $\Delta_{h,e}(T)$
near $T_c$ as
\begin{eqnarray}
\Delta_h^2(T)  &\approx& \frac{2}{1+N_h /N_e} \Delta_{BCS}^2(T),
\\ \nonumber
\Delta_e^2(T)  &\approx&   \frac{2}{1+N_e /N_h} \Delta_{BCS}^2(T)
\end{eqnarray}
\noindent with $\Delta_{BCS}^2(T)=\pi^2 \frac{8}{7\zeta(3)}
T_c^2(1-T/T_c)$. Combining the results of Eq.(4) and (6),
Eq.(1) provides
\begin{equation}
\frac{\Delta C}{T_c} \approx 4 \times (3.06)^2 N_{tot} \cdot(\bar{N}_h \bar{N}_e) .
\end{equation}
This is our key result. In contrast to the one band BCS
superconductor, Eq.(7) clearly shows that $\Delta C / T_c$ can
have a strong $T_c$ dependence through $\bar{N}_h \bar{N}_e$ even with a constant $N_{tot}$ (see Eq.(5)).
With doping in a given FePn/Ch
compound, $\bar{N}_{h}$ and $\bar{N}_{e}
(=1-\bar{N}_{h})$ varies over the range of $[0,1]$ of
\cite{doping}. As such if $(\bar{N}_h \bar{N}_e) \sim T_c^2$ for
some region of $\bar{N}_{h,e}$, we would obtain the BNC scaling.

Having analyzed the ideal case ($V_{intra}=0$), we numerically study
the more realistic cases, including the impurity
scattering effect. We solve the coupled gap equations
Eq.(2) for $\Delta_{h,e}(T)$ near $T_c$ and directly calculate
$\Delta C$ using Eq.(1). We find that the kinematic constraint of
the two band pairing model discovered above is robust. However in
order to explain the ideal BNC scaling $\Delta C / T_c \propto
T_c ^2$ in Ba(Fe$_{1-x}$TM$_x$)$_2$A$_2$ (TM=Co,Ni) as well as its strong deviation in
Ba$_{1-x}$K$_x$Fe$_2$A$_2$\cite{Calfield_K122}, we find that
the non-pair-breaking impurity scattering plays a crucial role.

\begin{figure}
\noindent
\includegraphics[width=90mm]{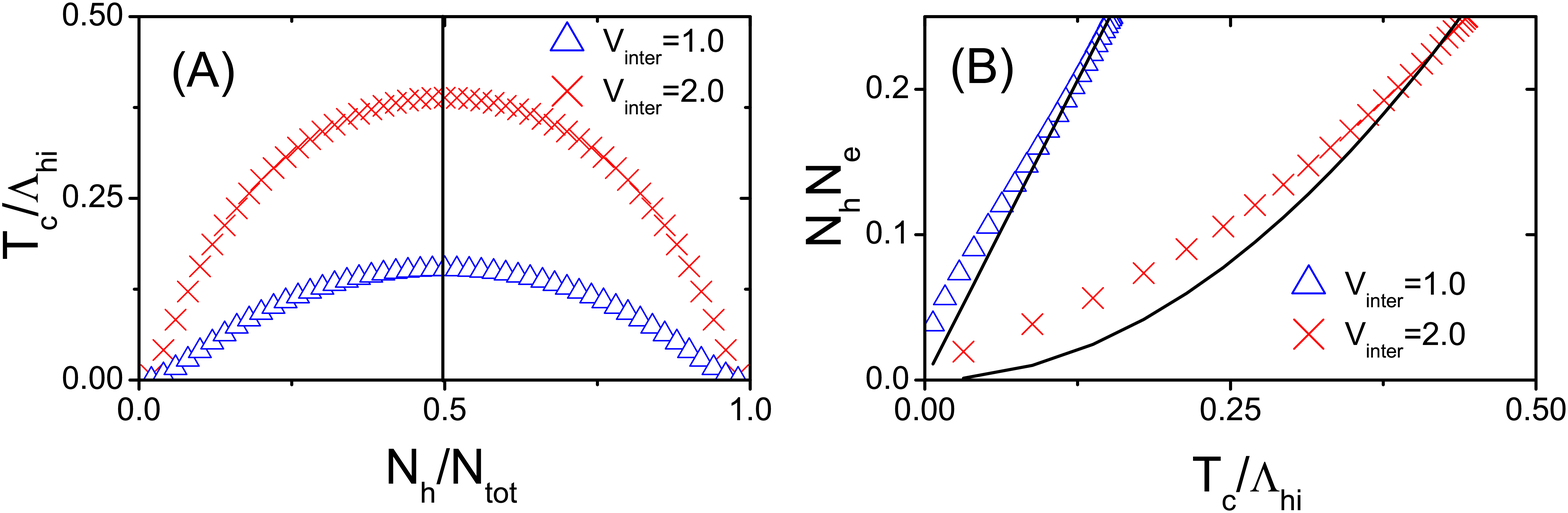}
\includegraphics[width=90mm]{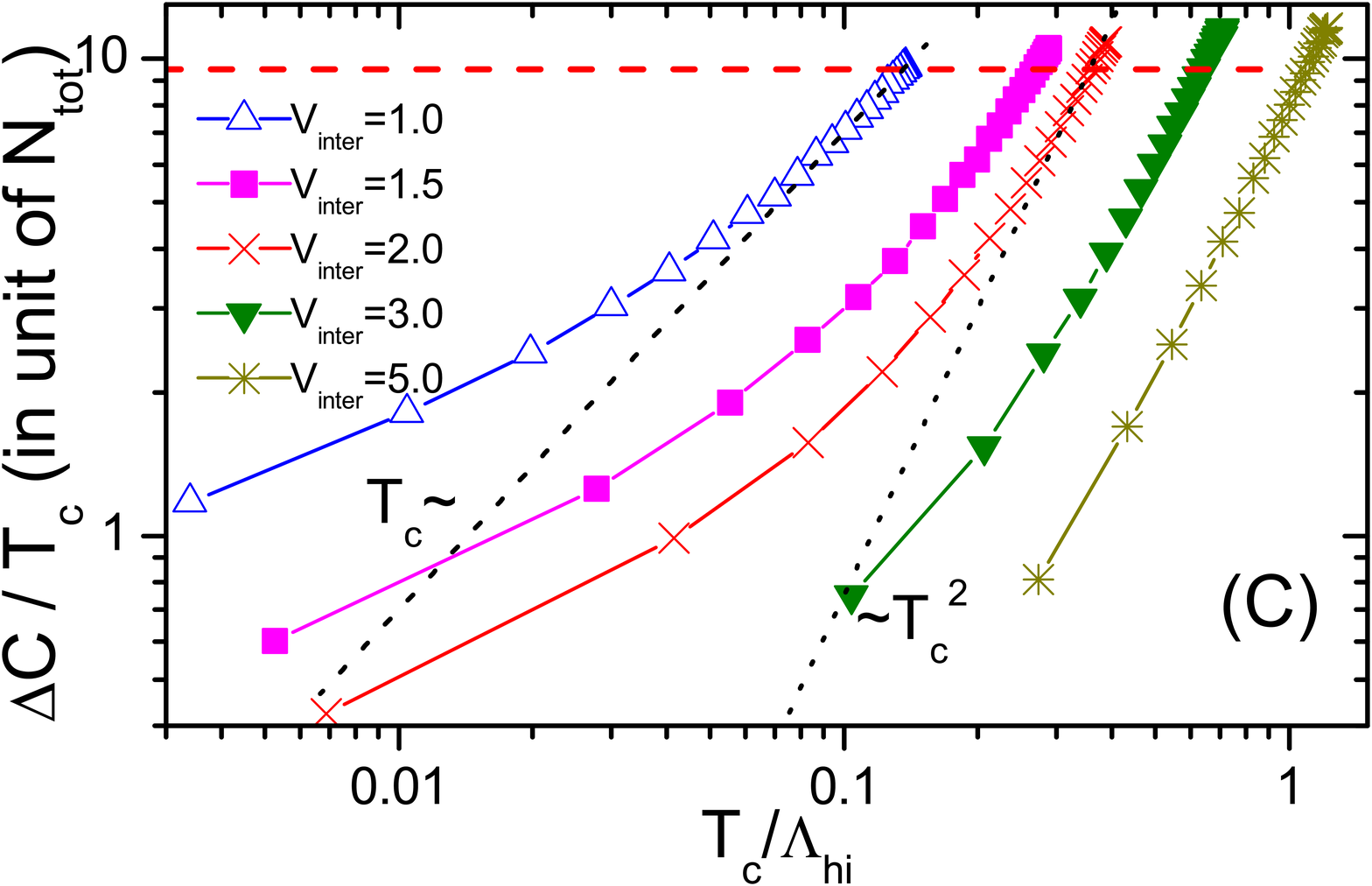}
\vspace{-1cm}
\caption{(Color online) (A) Numerical calculations of $T_c$ vs. $\bar{N}_h$ of the two band model for $\bar{V}_{inter}=1.0,$ and 2.0, respectively, with $\bar{V}_{intra}=0.0$ for both cases.
(B) Plots of $\bar{N}_h \bar{N}_e$ vs. $T_c$ with the same data of (A). Solid lines are
of $\sim T_c$ and $\sim T_c^2$, respectively.
(C) Numerical calculations of $\Delta C /T_c$ vs. $T_c$ for $\bar{V}_{inter}=1.0, 1.5, 2.0,$ and $3.0$, respectively,
with $\bar{V}_{intra}=0.0$ for all cases. Horizontal dashed line is the BCS limit of $9.36 N_{tot}$ and the dotted
lines of $\sim T_c$ and $\sim T_c^2$ (BNC scaling) are guides for the eyes.
\label{fig2}}
\end{figure}

{\it Numerical results.---} In Fig.1(A), we calculated $T_c$ vs. $\bar{N}_h$ of the
two band model Eq.(2) for $\bar{V}_{inter}=1.0$ and 2.0,
respectively, with $\bar{V}_{intra}=0$ for both cases.
Indeed, the calculated $T_c$ shows a strong dependence on $\bar{N}_h$,
symmetric with respect to $\bar{N}_h =0.5$ because $\bar{N}_h + \bar{N}_e =1$.
We plot the same data as $\bar{N}_h \cdot \bar{N}_e$ vs. $T_c$ in Fig.1(B).
In the case of $\bar{V}_{inter}=2.0$, we find $\bar{N}_h \cdot \bar{N}_e
\sim T_c ^2$ near the maximum $T_c$ region which is the necessary
condition for the BNC scaling from Eq.(7). It also shows that
the overall power of the relation $\bar{N}_h \cdot \bar{N}_e \sim T_c
^{\alpha}$ becomes weaker with the weaker pairing potential
$\bar{V}_{inter}$.

Now we calculate $\Delta C (\bar{N}_h)$ from Eq.(1) and Eq.(2),
and $\Delta C (\bar{N}_h)$ and $T_c(\bar{N}_h)$ are implicitly related through $\bar{N}_h \in
[0,1]$. In Fig.1(C), we plot $\Delta C / T_c$ vs.  $T_c$ in
log-log scale, for different pairing potentials
$\bar{V}_{inter}=1.0, 1.5, 2.0$ and 3.0, respectively, with
$\bar{V}_{intra}=0.0$ for all cases.  As hinted from Fig.1(B), we
can see the trend that the region of the BNC scaling $\Delta C /
T_c \sim T_c^2$ becomes widened near the maximum $T_c$ region with
increasing the pairing potential strength $\bar{V}_{inter}$. With
extensive numerical experiments, we found: (1) $\Delta C / T_c$
can become $\sim T_c^2$ for the whole region if $\bar{V}_{inter}
> 5.0$, but this strength of pairing potential is unrealistically large.
(2) Including $V_{intra} \neq 0.0$ does not change the general
behavior shown in Fig.1(C) as long as $\bar{V}_{intra}  < \bar{V}_{inter}/2 $.

While we have found that the BNC scaling can be realized in a
region near the maximum $T_c$ with the generic two
band model, we still need an extra mechanism to enhance the BNC scaling
for the wider region of $T_c$.
As shown in Fig.1(A) and Fig.1(B), $T_c$ is maximum when $\bar{N}_e =
\bar{N}_h =0.5$ and it quickly decreases with doping as $\bar{N}_h \bar{N}_e
\ll 0.25$ and accordingly one of the OPs, either $\Delta_h$
or $\Delta_e$, becomes tiny. Hence, the effect of impurity scattering on the tiny gap
becomes increasingly stronger for the lower $T_c$ region where the ratio $\bar{N}_e /\bar{N}_h$ is far from 1.
We found that this doping-dependent, therefore $T_c$-dependent, impurity effect changes the generic $\Delta C
/T_c$ vs. $T_c$ relation to a steeper relation at the lower $T_c$
region, hence enhances the region of the BNC scaling
even with a moderate strength of $\bar{V}_{inter}$.

Phenomenologically we introduce two parameters of the impurity
scattering in the two band model: $\Gamma_0$ (intra-band
scattering) and $\Gamma_{\pi}$ (inter-band scattering). As we
assumed the $\pm S$-wave state,
$\Gamma_{\pi}$ causes strong pair-breaking effect (e.g. suppression of $T_c$
and reduction of $\Delta_{h,e}$), but $\Gamma_0$
doesn't affect the superconductivity itself\cite{AG}. However, the
quasiparticle broadening is governed by the sum
$\Gamma_{tot}=\Gamma_0+\Gamma_{\pi}$ and the calculations of
$\Delta C$ from Eq.(1) should be generalized with this broadening of the quasiparticle spectra
as follows\cite{Skalski},

\begin{equation}
\Delta C = \sum_{i=h,e} N_i \Big( \frac{-d \Delta^2 _i}{d T} \Big) \Big|_{T_c}
\int_{0} ^{\infty} \frac{dx}{2}\Big[\frac{1}{\cosh^2(\frac{x}{2})}\Big] \frac{x^2}{x^2+(\frac{\Gamma_{tot}}{T_c})^2}
\end{equation}

\noindent where $x=\omega/T_c$. The standard pair-breaking effect
of $\Gamma_{\pi}$ still enters the pair-susceptibility
$\chi_{h,e}(T_c) = T_c \sum_{n} \int _{-\Lambda_{hi}} ^{\Lambda_{hi}} d\xi
\frac{1}{\tilde{\omega}_n ^2 +\xi^2}$ instead of Eq.(3) with
$\tilde{\omega}_n = \omega_n + \Gamma_{\pi}$, therefore $\Gamma_{\pi}$ directly affects $T_c$ and
$\Big( \frac{-d \Delta^2 _i}{d T} \Big)$ in Eq.(8). However,
increasing $\Gamma_{\pi}$ only\cite{Kogan2009} doesn't help for
producing the BNC scaling as discussed in the
Introduction\cite{Kogan note}. On the other hand, Eq.(8) above
shows that the total quasiparticle damping rate $\Gamma_{tot}$
entering the thermodynamic average part in Eq.(8) is more
important to determine $\Delta C$ vs. $T_c$.

\begin{figure}
\noindent
\includegraphics[width=90mm]{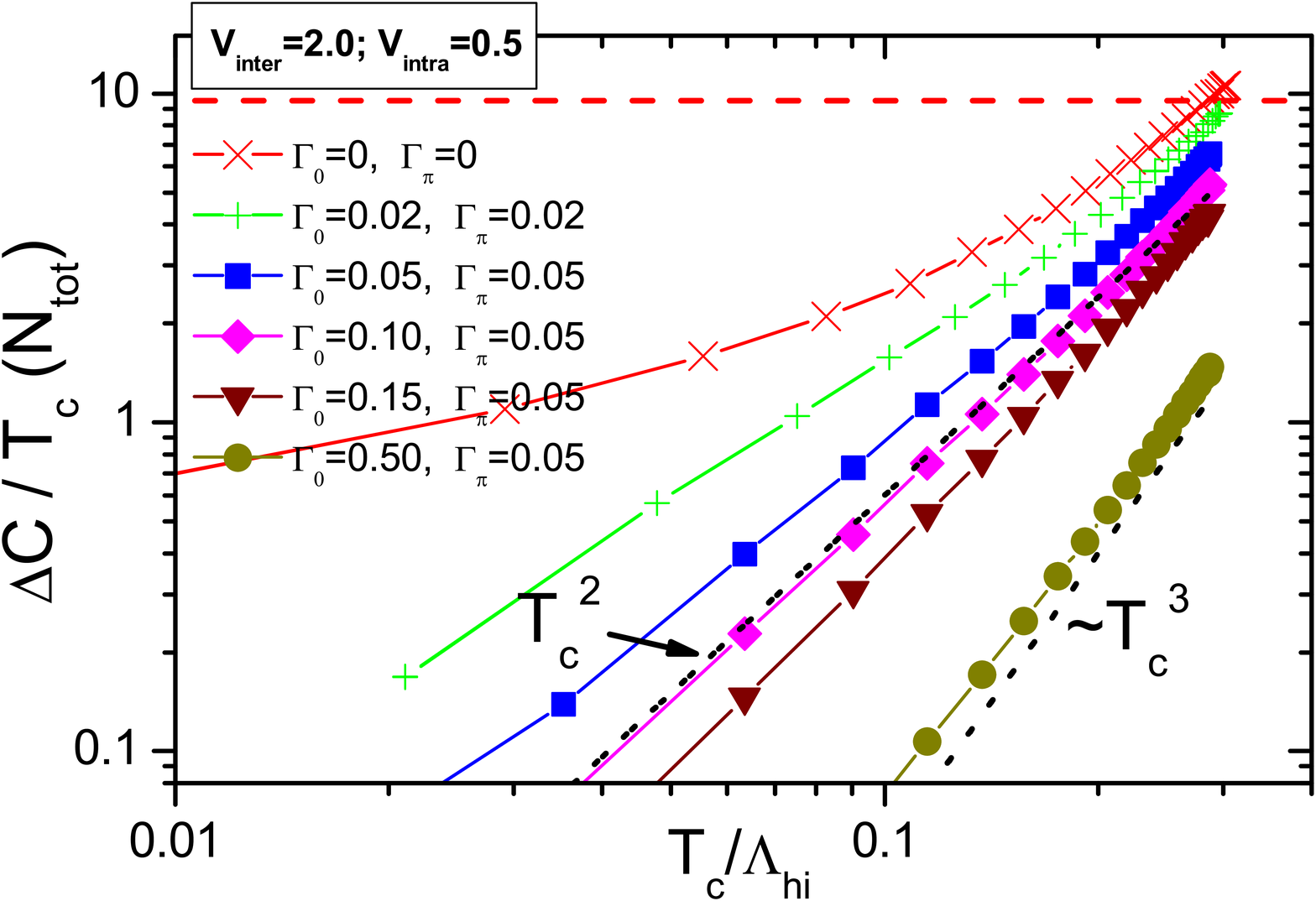}
\caption{(Color online) Numerical calculations of $\Delta C /T_c$ vs. $T_c$ with $\bar{V}_{inter}= 2.0$ and $\bar{V}_{intra}= 0.5$,
for different impurity scattering strengths of $\Gamma_0$ and $\Gamma_{\pi}$ (in unit of $\Lambda_{hi}$).
Horizontal dashed line is the
BCS limit of $9.36 N_{tot}$ and the dotted lines of $\sim T_c^2$  (BNC scaling)
and $\sim T_c^3$  (super-strong scaling) are guides for the eyes. \label{fig2}}
\end{figure}

In Fig.2, we show the numerical results of $\Delta C /T_c$ vs.
$T_c$ in log-log scale with a choice of a moderate strength of the
pairing potentials, $\bar{V}_{inter}=2.0$ and
$\bar{V}_{intra}=0.5$, and varied the impurity scattering rates
$\Gamma_0$, and $\Gamma_{\pi}$. Without impurity scattering (red
$"\times"$ symbols, $\Gamma_0$ = $\Gamma_{\pi}=0$), $\Delta C
/T_c$ shows the $T_c^2$ scaling only for the limited region
near the maximum $T_c$ and it quickly becomes flattened and slower
than $\sim T_c$. Interestingly, this behavior looks very similar
to the experimental data of Ba$_{1-x}$K$_x$Fe$_2$A$_2$
\cite{Calfield_K122}. Therefore, we speculate that the K-doping in
Ba$_{1-x}$K$_x$Fe$_2$A$_2$ doesn't introduce many impurity
scatterers. Next, only a small increase of impurities (green $"+"$
symbols, $\Gamma_0 =\Gamma_{\pi}=0.02$ in unit of $\Lambda_{hi}$)
immediately changes $\Delta C /T_c$ closer to $\sim T_c^2$ over
the whole $T_c$ range, and the case with $\Gamma_0 =0.1$ and
$\Gamma_{\pi}=0.05$ (pink $"\diamond"$ symbols) displays an ideal
BNC scaling $\Delta C /T_c \sim T_c^2$ for the entire range of
$T_c$. Finally, for demonstration purposes, we also show the case
with unrealistically large impurity scattering rates, $\Gamma_0 =0.5$ and
$\Gamma_{\pi}=0.05$ (dark green $"\circ"$ symbols), which displays
$\Delta C /T_c \sim T_c^3$, a super-strong scaling.

\begin{figure}
\noindent
\includegraphics[width=90mm]{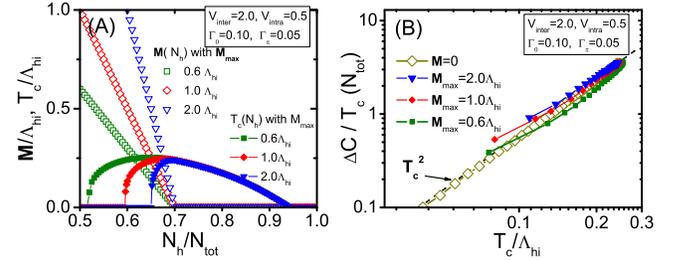}
\caption{(Color online) (A) Numerical calculations of $T_c$ vs. $\bar{N}_h$
with coexisting magnetic order $M(\bar{N}_h)$ of three different
strengths of $M_{max}=M(\bar{N}_h=0.5)$=0.6 (blue $\nabla$), 1.0 (red $\diamondsuit$),
and 2.0 (green box symbols) (in unit of $\Lambda_{hi}$).
The pairing interactions and the impurity scattering rates are chosen
$\bar{V}_{inter}=2.0$, $\bar{V}_{intra}=0.5$, and  $\Gamma_0 = 0.10$,
$\Gamma_{\pi}=0.05$, respectively.
(B) Calculated $\Delta C/T_c$ vs. $T_c$ for the corresponding
three cases of (A). The calculations of the region ($\bar{N}_h \in [0.7,1]$) where $M(\bar{N}_h)=0$
are all the same and displayed with dark yellow "$\diamondsuit$" symbols. \label{fig3}}
\end{figure}

{\it Coexistence region with magnetic and SC orders. ---}
Experiments showed that the BNC scaling continues to be valid
even when the spin density wave (SDW) order coexists with the SC order
in the underdoped regime. Now we would like to extend our model
including the magnetic order in the underdoped regime.
We took a simple phenomenological
approach ignoring the self-consistence between two OPs.
We consider only the hole doped region $\bar{N}_h \in [0.5,1]$,
because our model is symmetric with the hole and electron doping.
We arbitrarily chose the coexistence region for $0.5 \leq
\bar{N}_h <0.7$, just for the sake of demonstration, and then we
introduced the magnetic order $M(\bar{N}_h)$ for this region. The
magnetic OP $M(\bar{N}_h)$ linearly grows from zero at $\bar{N}_h
=0.7$ to a maximum value $M_{max}$ at $\bar{N}_h =0.5$ as shown in
Fig.3(A). When a finite $M$ exists, it affects the
superconductivity in two important ways: (1) it weakens the
SC pair susceptibility and we take the
simplest approximation as $\chi_{h,e}(T_c) = T_c \sum_{n} \int _0 ^{\Lambda_{hi}}d \xi
\frac{2}{\tilde{\omega}_n ^2 +\xi^2 + M^2}$ \cite{Vavilov2011}.
(2) The presence of SDW order $M$ also removes a part of the FSs.
Phenomenologically, we mimic this effect by linearly reducing the
total DOS $N_{tot}$ starting from $\bar{N}_h =0.7$ to a maximum
reduction at $\bar{N}_h =0.5$ as $N_{tot}(\bar{N}_h)=N_{tot}^0 [1-
a\frac{M(\bar{N}_h)}{\Lambda_{hi}}]$
($a=0.5$ was chosen for calculations in Fig.3). With these phenomenological Ans\"{a}tze,
we solved the $T_c$-equations from Eq.(2) with fixed pairing
interactions and damping ($\bar{V}_{inter}=2.0,
\bar{V}_{intra}=0.5$; and $\Gamma_0=0.10, \Gamma_{\pi}=0.05$) for
three different strengths of $M_{max}$ in Fig.3(A). The results
qualitatively simulate the experimental phase diagram: $T_c$
starts decreasing when $M(\bar{N}_h)$ starts developing from
$\bar{N}_h=0.7$ and the reduction of $T_c$ is faster with larger
magnetic order.

In Fig.3(B), $\Delta C /T_c$ vs. $T_c$ is calculated for the
corresponding three cases of Fig.3(A). The case of $M=0$ (dark yellow
"$\diamondsuit$" symbols), displaying
the $T_c ^2$ BNC scaling, is the same calculation as in Fig.2 with
$\Gamma_{0}=0.10$ and $\Gamma_{\pi}=0.05$ but only over $\bar{N}_h \in [0.7,1]$. Then the three other
solid symbols are the calculation results for the region of
$\bar{N}_h \in [0.5,0.7]$ with three different strengths of
magnetic order $M(\bar{N}_h)$ of Fig.3(A). The results of Fig.3(B) reveal an interesting behavior;
namely, although it is more natural to expect that
$\Delta C /T_c$ vs. $T_c$ with a coexisting magnetic order should behave differently from the one
without a magnetic order\cite{Vavilov2011}, the calculations of
Fig.3(B) with a crude phenomenological treatment of the coexisting
magnetic and SC orders show that it is quite robust to follow the
BNC scaling even with widely different strengths of $M$.
We trace the origin of this surprising
result to the fact that the underdoped region (i.e. where
$\bar{N}_h$ is near 0.5 and $T_c$ is maximum), when the magnetic
order is absent, is the region where the BNC scaling is best
obeyed due to the kinematic constraint of the multiband
superconductor (see Fig.1(c) and Fig.2). Therefore, even if the
magnetic order modifies the pair susceptibility
$\chi_{h,e}(M)$ and cuts out a part of DOS from $N_{tot}^0$, the
generic kinematic constraint of the multiband superconductor
dominated by $\bar{V}_{inter}$ is still operative.

{\it Summary and Conclusions ---} We showed that the puzzling BNC
scaling relation $\Delta C/T_c \sim T_c^2$\cite{BNC} observed in a
wide range of the FePn/Ch SC
compounds\cite{BNC,JSKim2011,Hardy,Gofryk2010,Greg2012,AsP1,AsP2,Calfield_Na122,HHWen}
is a manifestation of the generic property of the multiband
superconductor paired by a dominant inter-band pairing potential
$V_{inter} > V_{intra}$. The underlying mechanism is the kinematic
constraint $\frac{\Delta_h}{\Delta_e} \sim \sqrt{\frac{N_e}{N_h}}$
near $T_c$, and the subsequent
relations of $\Delta C \sim N_h N_e$ and $T_c(\sqrt{N_h N_e})$.
A consideration of the non-pair-breaking impurity effect which broadens the
quasiparticle spectra near $T_c$ also explains the evolution
from the ideal BNC scaling to its strong deviation as found in
Ba$_{1-x}$K$_x$Fe$_2$As$_2$\cite{Calfield_K122}.

{\it Acknowledgement -- } YB was supported by Grants No.
2013-R1A1A2-057535 funded by the National Research Foundation of
Korea. GRS was supported by the US Department of
Energy, contract no. DE-FG02-86ER45268.


\begin{references}
\bibitem{Kogan note}
In the strong pair-breaking limit ($\Gamma_{\pi} > T_c$) where
$\Delta F \propto - N(0)\frac{\Delta^4}{\Gamma_{\pi}^2}$ is valid,
the behavior of $\Delta(T)$ should also change to $\Delta^2(T)
\propto \Gamma_{\pi}^2 (1 - \frac{T}{T_c})$ instead of the BCS
behavior $\Delta^2(T) \propto T_c ^2 (1 - \frac{T}{T_c})$. Then
the SH jump in this limit becomes $\Delta C /T_c \sim
N(0)\frac{\Gamma_{\pi}^2}{T_c ^2}$, quite opposite to the BNC
scaling.
\bibitem{doping}
For the undoped parent compound BaFe$_2$As$_2$, our model assumes $\bar{N}_h =
\bar{N}_e=0.5$. For electron doping in
Ba(Fe$_{1-x}$Co$_x$)$_2$As$_2$, hole pocket(s) disappears
($\bar{N}_h=0$) around $x \approx 0.15$. For the hole doping cases, electron pockets(s) disappears ($\bar{N}_e=0$) in (Ba$_{1-x}$K$_x$)Fe$_2$As$_2$ when $x > 0.7$ but in
(Ba$_{1-x}$Na$_x$)Fe$_2$As$_2$ the electron pocket still remains
when $x \approx 1$. Therefore, the actual doping $"x"$ and our model doping parameter $\bar{N}_{h,e}$ should be properly rescaled according to the $T_c$ variation vs. doping.
\end{references}
\end{document}